\documentclass[12pt]{article}
\usepackage{epsf}
\usepackage{graphicx}
\usepackage{eepic}
\usepackage{color}
\usepackage{latexsym,epsfig}
\usepackage{subfigure}
\newcommand{\be}{\begin{equation}}
\newcommand{\ee}{\end{equation}}
\newcommand{\ba}{\begin{array}{c}}
\newcommand{\ea}{\end{array}}
\newcommand{\bqa}{\begin{eqnarray}}
\newcommand{\eqa}{\end{eqnarray}}

\begin{document}

\newcommand{\supercite}[1]{\textsuperscript{\cite{#1}}}
%\frenchspacing \linespread{1.2}
\begin{center}
{\Large\bf Pole analysis on unitarized $SU(3)\times SU(3)$ one loop
$\chi$PT amplitudes }
\\[15mm]
{\sc  Ling-Yun~Dai\footnote{e-mail: daily03@pku.edu.cn},
X.~G.~Wang~\footnote{e-mail: wangxuangong@pku.edu.cn. Address after
September 1,2011, Institute of High Energy Physics, Chinese Academy
of Science, Beijing 100049 }and H.~Q.~Zheng\footnote{e-mail:
zhenghq@pku.edu.cn}}
\\[5mm]
{\it   Department of Physics and State Key Laboratory of Nuclear
Physics and Technology, Peking University, Beijing 100871,
P.~R.~China}
\\[5mm]
\today

\begin{abstract}
We analyze $\pi\pi-K\bar{K}$ and $\pi\eta-K\bar{K}$ couple channel
 [1,1] matrix Pad\'e
amplitudes   of $SU(3)\times SU(3)$ chiral perturbation theory. By
fitting  phase shift and inelasticity data, we determine pole
positions in different channels ($f_0(980)$, $a_0(980)$,$f_0(600)$,
$K_0^*(800)$, $K^*(892)$, $\rho(770)$) and trace  their $N_c$
trajectories. We stress that a couple channel Breit--Wigner
resonance should exhibit two poles on different Riemann sheets that
reach the same position on the real axis when $N_c=\infty$. 
Poles are hence
classified using this criteria and we conclude that $K^*(892)$ and
$\rho(770)$ are unambiguous Breit--Wigner resonances.  For  scalars
the situation is much less clear. We find that $f_0(980)$ is a
molecular state rather than  a Breit--Wigner resonance, while
$a_0(980)$, though behaves oddly when varying $N_c$, does maintain a
twin pole structure.
\end{abstract}
\end{center}

\vspace{1cm}

\section{Introduction}
Chiral perturbation theory~\cite{Weinberg79,Gasser84,Gasser85}
 has been proven very successful in studying low energy hadron physics.  Nevertheless,
 when the energy goes higher, up to roughly 500MeV, chiral expansion breaks down since it's an expansion
 in terms of pseudo-goldstone boson mass and external momentum.
 Much efforts have also been made in exploring physics  beyond the validity
 domain of chiral expansions.
% including  resonance chiral perturbation theory
%(RChPT)\supercite{Ecker1,Ecker2}, unitarization
%approximation~\supercite{OOP98,Oller99, Oller97,Nieves99}, etc..
However, due to the non-perturbative nature it is very difficult to
study the problem in a model independent way.

When studying physics at higher energies,  one of the central
question was whether there exists (the lightest) scalar mesons such
as $f_0(600)$ (also called as $\sigma$) and $K_0^*(800)$  (also
called as $\kappa$). Dispersion technique played a crucial role in
firmly establishing the very existence of these resonances and in
determining their precise pole locations.~\cite{Zheng01}-\cite{KPY}
However,  the property of these scalar resonances, including that of
$f_0(980)$ and $a_0(980)$, remains mysterious, e.g.,
 whether they are
$q\bar{q}$ states, or $qq\bar{q}\bar{q}$ states or molecular states?
and what is the role they play in spontaneous chiral symmetry
breaking, and even in confinement? When studying  the substructure
of these scalar mesons, some authors try to investigate the two
photon decay width of the resonant
states~\cite{Pennington06}-\cite{MNW2}. One may also get helpful
information from the large $N_c$ behavior of the resonance
poles~\cite{Pelaez03}-\cite{Guo11}.
 The latter depends on the use of  the unitarized
  chiral perturbative amplitudes.   It is however known that Pad\'e approximants encounter
  severe difficulties, e.g., it violates crossing symmetry; it
  introduces spurious poles  on the physical sheet (and on un-physical
  sheets as well) and hence violates analyticity. Besides these
  deficiencies, unitarization approaches make correct predictions in some
  aspects, including the existence of $f_0(600)$ and $K^*_0(800)$
  resonances.

  This paper also devotes to the study of light hadron physics using
  unitarization approach of $SU(3)\times SU(3)$ $\chi$PT.
   Classification of resonance poles is an important topic in hadron
 physics.
  For this reason we point out that a couple channel
Breit--Wigner resonance should exhibit two poles on different
Riemann sheets and reach the same position on the real axis when
$N_c=\infty$. This simple criteria is used to classify poles that
emerge in the unitarized amplitudes. It is clearly seen that
$K^*(892)$ and $\rho(770)$ are unambiguous Breit--Wigner resonances,
which, in our understanding, is what Pad\'e approximants can predict
reliably. This observation encourages us to use this criteria to
analyze scalar resonances and we conclude that $f_0(980)$ is
$\textbf{not}$ a Breit--Wigner resonance, it is most likely a $K\bar
K$ (virtual) bound state when $N_c$ is large. On the other side,
$a_0(980)$ though behaves oddly when varying $N_c$, does maintain a
twin pole structure. The property of $f_0(600)$ and $K^*_0(800)$ is
also reinvestigated and previous results obtained in the single case
is confirmed.

This paper is organized as follows: This section is the
introduction, in section 2 we give a brief introduction to the
perturbation calculation of $\pi\pi\rightarrow K\bar{K}$,
$\pi\pi\rightarrow \pi\pi$, $K\bar{K}\rightarrow K\bar{K}$,
$\pi\eta\rightarrow \pi\eta$, and $\pi\eta\rightarrow K\bar{K}$
scattering amplitudes as well as partial wave projection and Pad\'e
unitarization.
%\supercite{Meissner91,Pelaez02}
 In section 3
We fit the experimental phase shift and inelasticity  to fix the
LECs and extract pole positions. Section 4 devotes to the study of
the property of poles by analyzing their $N_c$ trajectories. We make
physical discussion and conclusion in section 6, including a
comparison between our results and previous results found in the
literature.

%%%%%%%%%%%%%%%%%%%%%%%%%%%%%%%%%%%%%%%%%%%%%%%%%%%%%%%%%%%%%%%%%%%%%%%%%%%%%%%%%%%%%%%%%%%%%%%%%%%%%%%%%%%%%%%%%%
\section{Scattering amplitudes up to $\mathcal
{O}(p^4)$ and partial wave projections}\label{ChPTamp}
\subsection{Effective Lagrangian} At lowest order, the chiral
lagrangian at $\mathcal {O}(p^2)$ is
\begin{equation}\label{l2}
\mathcal
{L}_2=\frac{f_0^2}{4}<\partial_{\mu}U^{\dag}\partial^{\mu}U+\mathcal
{M}(U+U^{\dag})>\ ,
\end{equation}
where $<>$ stands for the trace of the $3\times 3$ matrices built
from $U(\Phi)$ and $\mathcal{M}$,
\begin{equation}
U(\Phi)=\exp(\frac{i\sqrt{2}}{f_0}\Phi)\ ,
\end{equation}
where $\Phi$ is expressed in terms of the Goldstone boson fields as
\begin{equation}
\Phi(x)=\left(
          \begin{array}{ccc}
            \frac{1}{\sqrt{2}}\pi^0+\frac{1}{\sqrt{6}}\eta & \pi^+ & K^+ \\
            \pi^- & -\frac{1}{\sqrt{2}}\pi^0+\frac{1}{\sqrt{6}}\eta & K^0 \\
            K^- & \bar{K}^0 & -\frac{2}{\sqrt{6}}\eta \\
          \end{array}
        \right)\ .
\end{equation}
The mass matrix $\mathcal{M}$ is given by
\begin{equation}
\mathcal{M}=\left(
              \begin{array}{ccc}
                \hat{m}_{\pi}^2 & 0 & 0 \\
                0 & \hat{m}_{\pi}^2 & 0 \\
                0 & 0 & 2\hat{m}_{K}^2-\hat{m}_{\pi}^2 \\
              \end{array}
            \right)
\end{equation}
in the isospin limit, where $\hat{m}$ means bare masses.\\
The chiral lagrangian of $\mathcal{O}(p^4)$ can be written
as,~\cite{Gasser85}
\begin{eqnarray}
\mathcal{L}_4&=&L_1 <\partial_{\mu}U^{\dag}\partial^{\mu}U>^2+
L_2<\partial_{\mu}U^{\dag}\partial_{\nu}U><\partial^{\mu}U^{\dag}\partial^{\nu}U>+\nonumber\\
&&
L_3<\partial_{\mu}U^{\dag}\partial^{\mu}U\partial_{\nu}U^{\dag}\partial^{\nu}U>+
L_4<\partial_{\mu}U^{\dag}\partial^{\mu}U><U^{\dag}\mathcal{M}+\mathcal{M}^{\dag}U>+\nonumber\\
&& L_5
<\partial_{\mu}U^{\dag}\partial^{\mu}U(U^{\dag}\mathcal{M}+\mathcal{M}^{\dag}U)>+L_6
<U^{\dag}\mathcal{M}+\mathcal{M}^{\dag}U>^2+\nonumber\\
&& L_7 <U^{\dag}\mathcal{M}-\mathcal{M}^{\dag}U>^2+
L_8<\mathcal{M}^{\dag}U\mathcal{M}^{\dag}U+U^{\dag}\mathcal{M}U^{\dag}\mathcal{M}>
\ .
\end{eqnarray}
%The renormalized coupling constants are defined by
%\begin{equation}
%L_i=L_i^r(\mu)+\Gamma_i\frac{\mu^{d-4}}{(4\pi)^2}\{\frac{1}{d-4}-\frac{1}{2}[\ln
%4\pi+1+\Gamma'(1)]\}
%\end{equation}
%in dimensional regularization scheme, with an arbitrary scale $\mu$.
%The coefficients $\Gamma_i$ are given by~\cite{Gasser85}.
%%%%%%%%%%%%%%%%%%%%%%%%%%%%%%%%%%%%%%%%%%%%%%%%%%%%%%%%%%%%%%%%%%%%%%%%%%%%%%%%%%%%%%%%
\subsection{Amplitudes}
\begin{figure}[h]%
\begin{center}%
%\vspace{2cm}
\mbox{\epsfxsize=80mm\epsffile{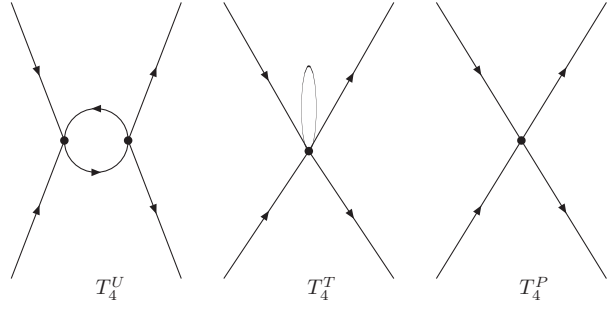}}%
%\vspace{-2cm}
\caption{\label{diagrams}Feynman diagrams of $\mathcal {O}(p^4)$}
\end{center}%
\end{figure}

\indent At $\mathcal {O}(p^4)$ one has to calculate the diagrams
shown in Fig.~\ref{diagrams}. $T_4^T$ represents contributions
coming from the $\mathcal {L}_2$ ChPT lagrangian with six fields and
a tadpole. $T_4^U$ represents the loops constructed from $\mathcal
{L}_2$ vertices with four fields. These loops include contributions
from s, t and u channels. $T_4^P$ amounts for the $\mathcal
{O}(p^4)$ polynomial contribution coming from $\mathcal {L}_4$
lagrangian.
In addition, one needs to take into account renormalization of
masses, decay constants and wave functions. The relations between
$m_P(P=\pi, K, \eta)$ and $\hat{m}$, and the ones between $f_P$ and
$f_0$ can be obtained from~\cite{Gasser85}. The scattering
amplitudes have been given in~\cite{Pelaez02}, all of which are
expressed in terms of $f_{\pi}$. As an exercise, we have
recalculated all the 1--loop amplitudes $\pi\pi\to\pi\pi$,
$\pi\pi\to \bar KK$, $\bar KK\to \bar KK$, $K\pi\to K\eta$ and
$\pi\eta\to\bar KK$ and confirm previous results found in the
literature.
%%%%%%%%%%%%%%%%%%%%%%%%%%%%%%%%%%%%%%%%%%%%%%%%%%%%%%%%%%%%%%%%%%%%%%%%%%%%%%%%%%%%%%%%%%%%%%%%%%%%%%%%%%%%%%%%%%%%%%%%%
%\section{Unitarization}
\subsection{Partial wave expansion}\label{Unitarization}
The iso-spin decomposed amplitudes
  for $\pi\pi\rightarrow\pi\pi$ scattering are
\begin{eqnarray}
T^{I=0}&=&T(\pi^0\pi^0\rightarrow\pi^0\pi^0)+2T(\pi^+\pi^-\rightarrow\pi^0\pi^0)\
,\nonumber\\
T^{I=1}&=&2T(\pi^+\pi^-\rightarrow\pi^+\pi^-)-T(\pi^+\pi^-\rightarrow\pi^0\pi^0)-T(\pi^0\pi^0\rightarrow\pi^0\pi^0)\ ,\nonumber\\
T^{I=2}&=&T(\pi^0\pi^0\rightarrow\pi^0\pi^0)-T(\pi^+\pi^-\rightarrow\pi^0\pi^0)\
;
\end{eqnarray}
for $\pi\pi\rightarrow KK$ are
\begin{eqnarray}
T^{I=0}(s,t,u)&=&\sqrt{6}T(\pi^0\pi^0\rightarrow K^+K^-)\ ,\nonumber\\
T^{I=1}(s,t,u)&=&2[T(\pi^+\pi^-\rightarrow
K^+K^-)-T(\pi^0\pi^0\rightarrow K^+K^-)]\ ;
\end{eqnarray}
for $KK\rightarrow KK$ scattering are
\begin{eqnarray}
T^{I=0}(s,t,u)&=&T(K^+K^-\rightarrow K^+K^-)+T(K^+K^-\rightarrow
K^0\bar{K}^0)\ ,\nonumber\\
T^{I=1}(s,t,u)&=&T(K^+K^-\rightarrow K^+K^-)-T(K^+K^-\rightarrow
K^0\bar{K}^0)\ ;
\end{eqnarray}
for $\pi\eta\rightarrow \pi\eta$ scattering are
\begin{equation}
T^{I=1}(s,t,u)=T(\pi^0\eta\rightarrow\pi^0\eta)\ ;\\
\end{equation}
and for $\pi\eta\rightarrow KK$ scattering are
\begin{equation}
T_{12}^{I=1}(s,t,u)=-\sqrt{2}T(\pi^0\eta\rightarrow K^+K^-)\ .
\end{equation}
The projection in definite angular momentum J is given by
\begin{equation}
T^{(I,J)}=\frac{1}{32N\pi}\int_{-1}^{1}d(cos\theta)T^I(s,t)P_J(cos\theta)\
,
\end{equation}
where N=${2}$ for elastic $\pi\pi$ and $\eta\eta$ scatterings,
and N=1 for other states. As an exercise we have re-calculated all
of the above amplitudes and confirm the previous results found in
the literature.

%%%%%%%%%%%%%%%%%%%%%%%%%%%%%%%%%%%%%%%%%%%%%%%%%%%%%%%%%%%%%%%%%%%%%%%%%%%%%
\subsection{Pad\'e approximation}
In two-channel case, the unitarity condition reads,
\begin{eqnarray}\label{unitarity}
\mathrm{Im} T_{11}&=&T_{11}\rho_1
T_{11}^{*}\theta(s-4m_{\pi}^2)+T_{12}\rho_2
T_{21}^{*}\theta(s-4m_{K}^2)\ ,\nonumber\\
\mathrm{Im} T_{12}&=&T_{11}\rho_1
T_{12}^{*}\theta(s-4m_{\pi}^2)+T_{12}\rho_2
T_{22}^{*}\theta(s-4m_{K}^2)\ ,\nonumber\\
\mathrm{Im} T_{22}&=&T_{21}\rho_1
T_{12}^{*}\theta(s-4m_{\pi}^2)+T_{22}\rho_2
T_{22}^{*}\theta(s-4m_{K}^2)\ ,
\end{eqnarray}
where the $IJ$ superscripts have been suppressed for simplicity.
$\rho_i=2q_i/\sqrt{s}$ is the phase space function, with $q_i$ the
center of mass momentum in the state $i$. Here, for I,J=0,0 and 1,1
channel, the subscript 1 represents for $\pi\pi$, 2 for $\bar KK$.
The third equation only holds true above $4m_K^2-4m_{\pi}^2$ along
the real axis rather than above $4m_{\pi}^2$~\cite{Kennedy62}.
%In the
%paper~\cite{Oller99}, a rather general scheme was derived to obtain
%final unitarized amplitudes from the $\mathcal{O}(p^2)$ and
%$\mathcal{O}(p^4)$ ChPT scattering amplitudes, which is equivalent
%to matrix $Pad\acute{e}$ approximation,
The [1,1] matrix Pad\'e approximant reads,
\begin{equation}
T=T^{(2)}\cdot[T^{(2)}-T^{(4)}]^{-1}\cdot T^{(2)}\ .
\end{equation}
However, it will cause the difficulty that the left hand cut
$(-\infty, 4m_K^2-4m_{\pi}^2]$ will appear not only in $T_{22}$, but
also in the other two amplitudes as well. This difficulty also
happens in couple channel K-Matrix parametrization~\cite{Benitez78}.
Hence, Eq.~(\ref{unitarity}) is satisfied exactly only above $\bar
KK$ threshold, although the deviation from unitarity may be
numerically small in some cases~\cite{Oller99}.

Partial wave $S$ matrix elements are given by
\begin{eqnarray}
S_{11}&=&1+2i\rho_1(s)T_{11}(s)\ ,\nonumber\\
S_{12}&=&2i\sqrt{\rho_{1}(s)\rho_2(s)}T_{12}(s)\,\nonumber\\
S_{22}&=&1+2i\rho_2(s)T_{22}(s)\ .
\end{eqnarray}
 In the physical region above the second threshold, the phase
shifts and inelasticity can be related to T-matrix through the
well-known parametrization,
\begin{equation}
S=\left(
          \begin{array}{cc}
     \eta e^{2i\delta_1}                              & i\sqrt{1-\eta^2}e^{i(\delta_1+\delta_2)} \\
     i\sqrt{1-\eta^2}e^{i(\delta_1+\delta_2)} & \eta e^{2i\delta_2}                              \\
          \end{array}
        \right)\ .
\end{equation}

When discussing $\pi\pi$ scatterings,
there have been many discussions in the literature on Pad\'e approximation and its variations.
This method is remarkable and is very helpful in some aspects, e.g., it builds a unitary scattering
amplitude naturally from  a perturbative one,
it correctly predicts the existence of $\sigma$,
$\kappa$ resonances at qualitative level, etc.. Nevertheless the unitarization method also maintains severe
deficiencies, e.g., it violates crossing symmetry and introduces spurious poles on the complex $s$-plane and hence
violates analyticity. Therefor we try to be careful when making conclusions on
outputs from a Pad\'e amplitudes.

%%%%%%%%%%%%%%%%%%%%%%%%%%%%%%%%%%%%%%%%%%%%%%%%%%%%%%%%%%%%%%%
\section{Fit results and $N_c$ trajectories}\label{results}
In this part we will present two types of fits. In Fit I we fit in
I,J=0,0 and I,J=1,1 $\pi\pi$, $\bar KK$ couple channel data. This
enables us to focus on physics of $f_0(600)$ and $f_0(980)$. In Fit
II we extend the above fit by including I,J=2,0 channel $\pi\pi$
elastic scattering data and I,J=1,0 $\pi\eta\to \bar KK$ data. The
latter is very important when exploring physics of $a_0(980)$.
\subsection{Fit 1}
\begin{figure}[h]%
\begin{center}%
%\vspace{2cm}
\subfigure[]{\label{a}\mbox{\epsfxsize=70mm\epsffile{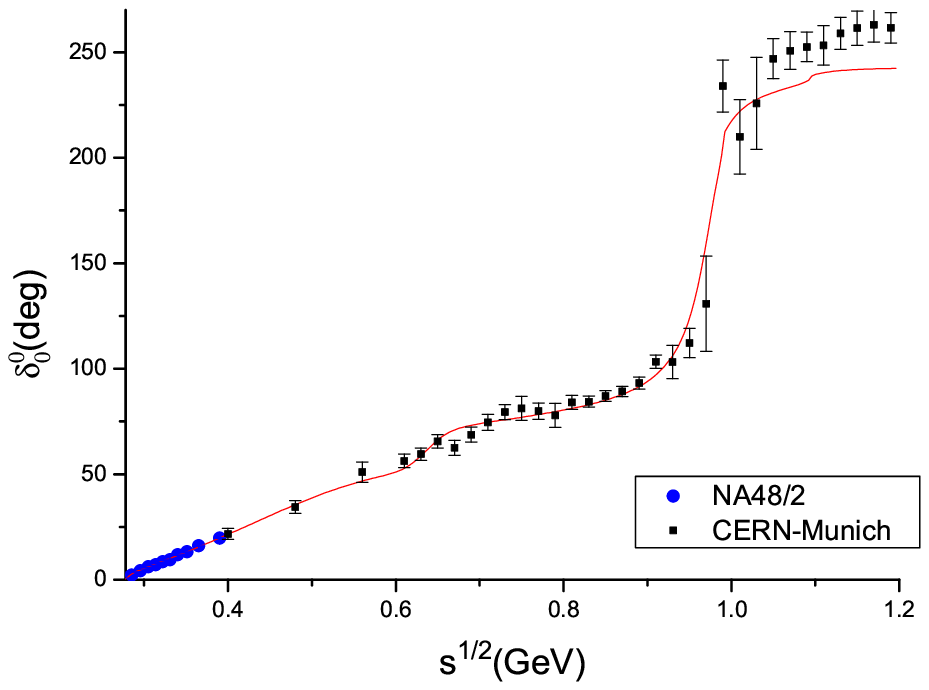}}}%
\subfigure[]{\label{b}\mbox{\epsfxsize=70mm\epsffile{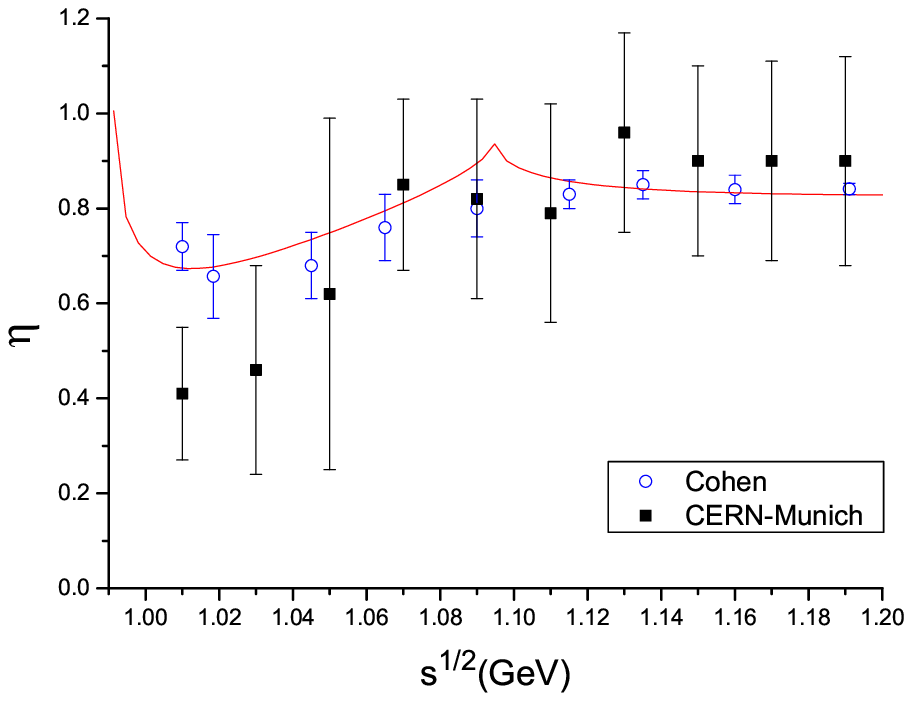}}}\\
\subfigure[]{\label{c}\mbox{\epsfxsize=70mm\epsffile{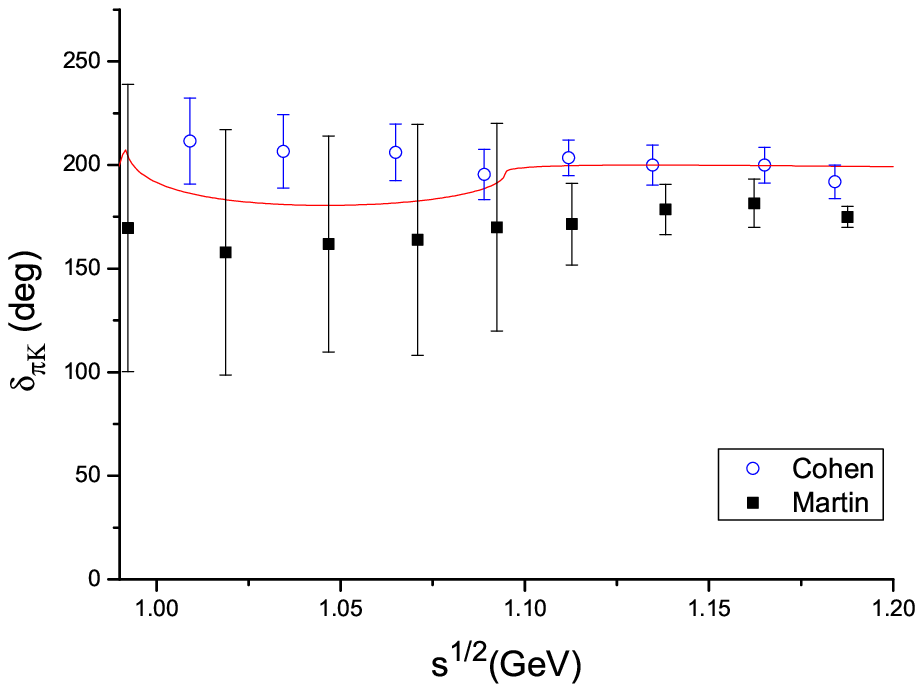}}}%
\subfigure[]{\label{d}\mbox{\epsfxsize=70mm\epsffile{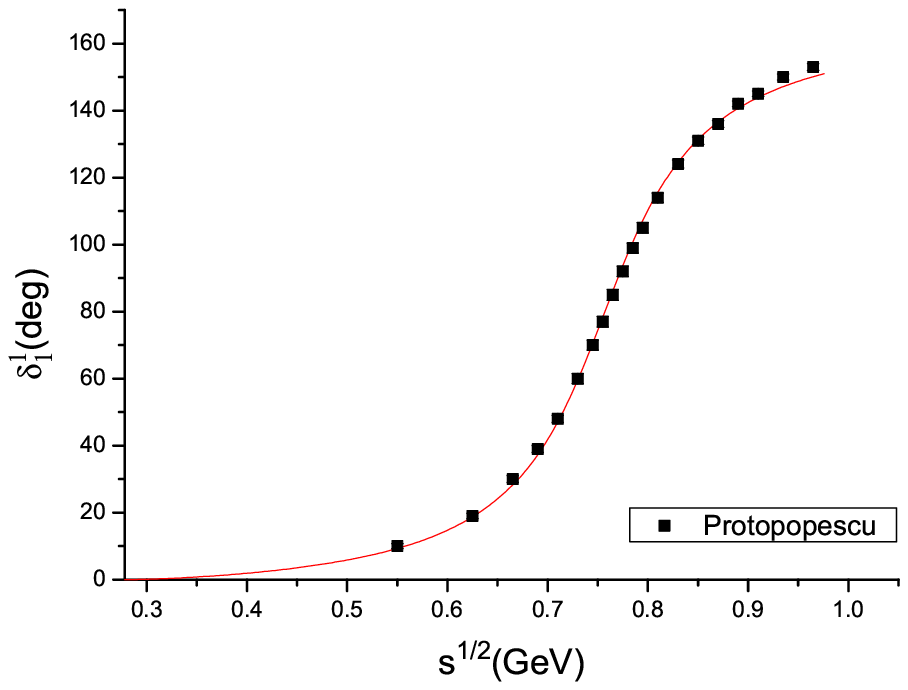}}}\\%
%\vspace{-2cm}
\caption{\label{fitresults} From Fit I: a) $I=0$ S wave $\pi\pi$
phase shift; b) inelasticity; c) $\delta_{\pi K}$; d) I,J=1,1
channel $\pi\pi$ phase shift.}
\end{center}%
\end{figure}
We first concentrate on $\pi\pi,\bar{K}K$ couple channel system and
the properties of $\sigma$ and $f_0(980)$. The data sets are taken
from~\cite{CERN-Munich}-\cite{Martin} for I,J=0,0 channel (fit range
from threshold to 1.2GeV) and~\cite{Protopopescu} for I,J=11 channel
(fit range from threshold to 1.0GeV). The fit results are shown in
Fig.~\ref{fitresults} with $\chi^2_{d.o.f}=186.3/95=1.96$, and the
LECs are given by Tab.~\ref{LECS}.
%%%%%%%%%%%%%%%%%%%%%%%%%%%%%%%%%%%%%%%%%%%%%%%%%%%%%%%%%%%%%%%%%%%%%%%%%%%%%%%%%%%%%%%%%%%
\begin{table}
\begin{center}
 \begin{tabular}  {|c|c|c|c|c|c|}\hline
 \   &            Fit\ I      &      Fit\ II             &       CHPT        &  Nc\ order \\  \hline
$L_1$      &  $0.98\pm 0.06$   &  $1.25\pm 0.02$           &  $0.40\pm 0.3$    & $\mathcal {O}(Nc)$ \\
$L_2$      &  $1.39\pm 0.06$   &  $1.63\pm 0.03$          &  $1.35\pm 0.3$    & $\mathcal {O}(Nc)$ \\
$L_3$      &  $-3.65\pm 0.11$  &  $-4.07\pm 0.02$           &  $-3.5\pm 1.1$    & $\mathcal {O}(Nc)$ \\
$L_4$      &  $0.40\pm 0.03$   &  $0.30\pm 0.02$          &  $-0.30\pm 0.5$   & $\mathcal {O}(1)$ \\
$L_5$      &  $0.21\pm 0.73$   &  $-1.44\pm0.42$           &  $1.40\pm 0.5$    & $\mathcal {O}(Nc)$ \\
$L_6$      &      \            &       \                  &
$-0.2\pm0.3$ & $\mathcal
{O}(1)$ \\
$L_7$      &      \           &       \                   &
$-0.40\pm 0.20$ & $\mathcal
{O}(1)$ \\
$L_8$      &      \           &       \                 & $0.90\pm
0.30$ & $\mathcal
{O}(Nc)$ \\
$2L_6+L_8$ &  $1.42\pm 0.41$  &  $0.32\pm0.13$            & \ &
 \  \\
$2L_7+L_8$ &       \          &  $1.38\pm0.38$            & \ &
 \  \\ \hline
 \end{tabular}
 \caption{\label{LECS}Low energy constants($\times 10^{-3}$) and $N_c$ dependence.}
 \end{center}
\end{table}
%%%%%%%%%%%%%%%%%%%%%%%%%%%%%%%%%%%%%%%%%%%%%%%%%%%%%%%%%%%%%%%%%%%%%%%%%%%%%%%%%%%%%%%%%%%%%
The $\pi\pi$ scattering length of I,J=0,0 channel is
\begin{equation}
a^0_0=0.213 m_{\pi}^{-1}.
\end{equation}
Pole positions of $\sigma(600)$ and $f_0(980)$ on the second sheet
of $\sqrt{s}$ plane are
\begin{equation}
\sqrt{s_{\sigma}}=(0.444-0.245i)\mathrm{GeV},\ \ \ \ \ \ \
\sqrt{s_{f_0}}=(0.974-0.026i)\mathrm{GeV},
\end{equation}
which are in agreement with recent determinations~\cite{CGL}-\cite{KPY}.\\
%%%%%%%%%%%%%%%%%%%%%%%%%%%%%%%%%%%%%%%%%%%%%%%%%%%%%%%%%%%%%%%%%%%%%%%%%%%%%%%%%%%%%%%%%%%%%%%%%%%%%%%%%%%%%

%%%%%%%%%%%%%%%%%%%%%%%%%%%%%%%%%%%%%%%%%%%%%%%%%%%%%%%%%%%%%%%%%%%%%%%%%%%%%%%%%%%%%%%%%%%%%%%%%%%%%%%%%%%%
\begin{figure}[h]%
\begin{center}%
%\vspace{2cm}
\mbox{\epsfxsize=70mm\epsffile{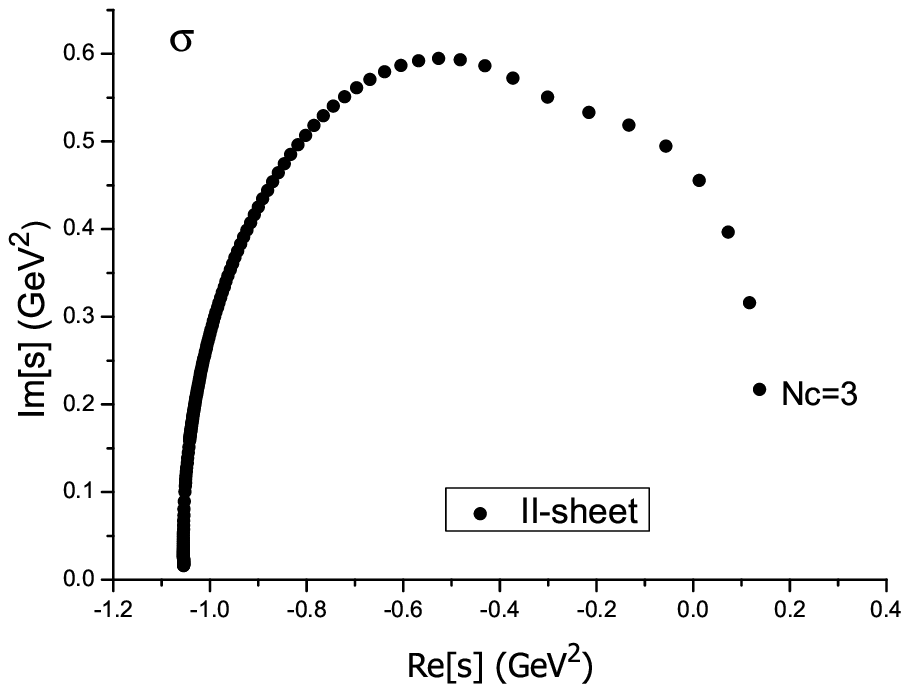}}%
\mbox{\epsfxsize=70mm\epsffile{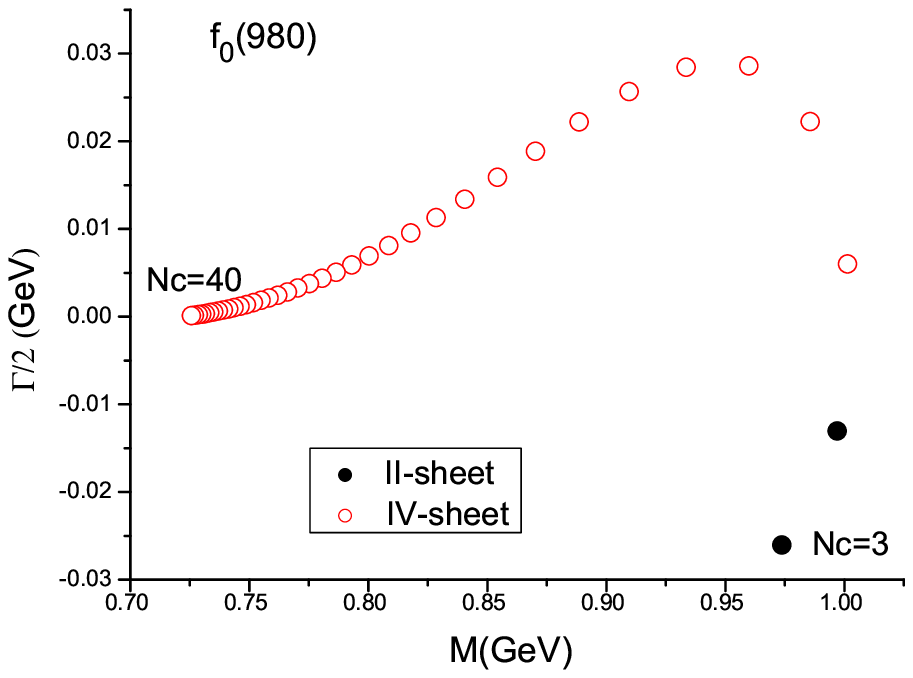}}
%\vspace{-2cm}
\caption{\label{NcTrajectorys}$N_c$ trajectories of $\sigma$(left)
and $f_0(980)$(right). Full circle: on sheet II; open circle: on
sheet IV}
\end{center}%
\end{figure}
We numerically trace the resonance pole positions when $N_c$ varies.
The $N_c$ trajectories of $\sigma$ and $f_0(980)$ are shown in
Fig.~\ref{NcTrajectorys}. We find that the $\sigma$ pole will
move towards negative real axis on the complex s plane when $N_c$ is
large enough.
This  result is previously  observed using single
channel [1,1] Pad\'e amplitude~\cite{Sun}.~\footnote{This phenomenon is also found
in \cite{Guo11}.}
 This odd trajectory
of $f_0(600)$ makes the topic  very puzzling and interesting, which
certainly deserves further investigations. Actually the pole's
destination in large $N_c$ and chiral limit can be determined in the
single channel approximation~\cite{Sun}: \be\label{sigmapole}
 s_\sigma={3f_\pi^2\over (44L_1+28L_2+22L_3)}\ .
\ee
In theory, this analytic solution
 also gives other two possibilities: fall
down to the positive real axis, or move to infinity, by slightly
tuning parameters. In the more complicated couple channel case,
analytic solution of $\sigma$ pole position is not available.
However we have performed a limited numerical tests by varying LECs
and did not find other possibilities except falling down to negative
real axis. Nevertheless, in [1,2] Pad\'e approximant it is found
that the $\sigma$ pole trajectory fall down on the positive real
axis at roughly $s_\sigma \sim 1GeV$~\cite{Sun}. So the only robust
conclusion one can make on the trajectory is that, unlike the $\rho$
trajectory, the scalar one does not like to fall down to its fateful
destination -- the real axis.

It is more important to stress that, it is actually understood well
what is the condition in obtaining Eq.~(\ref{sigmapole}), which
reveals partly the secret hidden in Pad\'e approximation. Using the
PKU dispersive representation one obtains the same result as
 Eq.~(\ref{sigmapole}) under two additional assumptions: 1) neglect
all crossed channel resonance exchanges (which contribute at leading
order in 1/$N_c$ expansion); 2) assume in $s$ channel there is only
one pole involves~\cite{Cillero}.

The trajectory of $f_0(980)$ is quite different.  The $f_0(980)$
pole moves into  upper half plane of sheet IV  from the lower half
of sheet II in the $\sqrt{s}$-plane, winding around the branch point
at the $K\bar{K}$ threshold. But, it will fall on the positive real axis, above
$\pi\pi$ threshold but below $\bar KK$ threshold on sheet IV
eventually. This observation is not found in previous literature. In
Ref.~\cite{Uehara04}, it is also noticed that $f_0(980)$ pole moves
from sheet II to sheet IV, but jump to a quick conclusion that the
pole moves to infinity without tracing the trajectory for lager
value of $N_c$. If our result is correct then it predicts the
following picture: when $N_c=3$ since $f_0(980)$ lies on sheet II,
it is  a $\bar KK$ bound state if $\pi\pi$ channel is switched off,
the destination of $f_0(980)$ indicates that it be a virtual bound
state of $\bar KK$ when $N_c$ becomes large. The conclusion is based
on Morgan's pole counting mechanism~\cite{morgan92}, since no other
poles near the $\bar KK$ threshold are found.\footnote{This
technique is also used to support that the X(3872) be mainly a $\bar
qq$ state~\cite{Zhangou}.}
%%%%%%%%%%%%%%%%%%%%%%%%%%%%%%%%%%%%%%%%%%%%%%%%%%%%%%%%%%%%%%%%%%%%%%%%%%%%%%%%%%%%%%%%%%%%%%%%%%%%%%%%%%%
\subsection{Fit 2}
\begin{figure}[h]%
\begin{minipage}[h]{0.5\textwidth}
\centering \subfigure[ ]{ \label{a}
\includegraphics[width=0.9\textwidth]{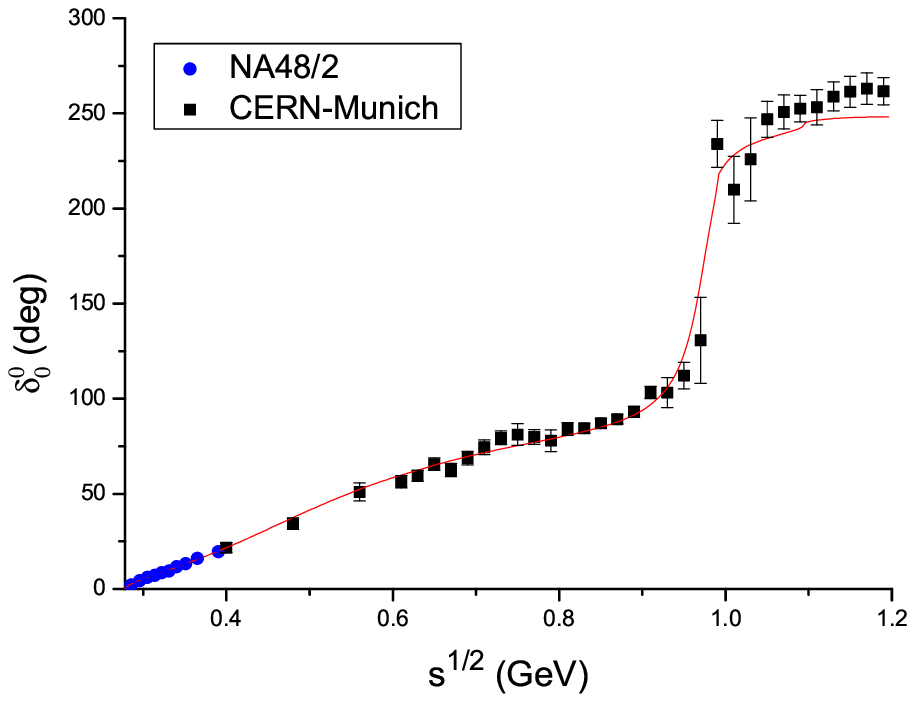}}%
\end{minipage}
\begin{minipage}[h]{0.5\textwidth}
\centering \subfigure[ ]{ \label{b}
\includegraphics[width=0.9\textwidth]{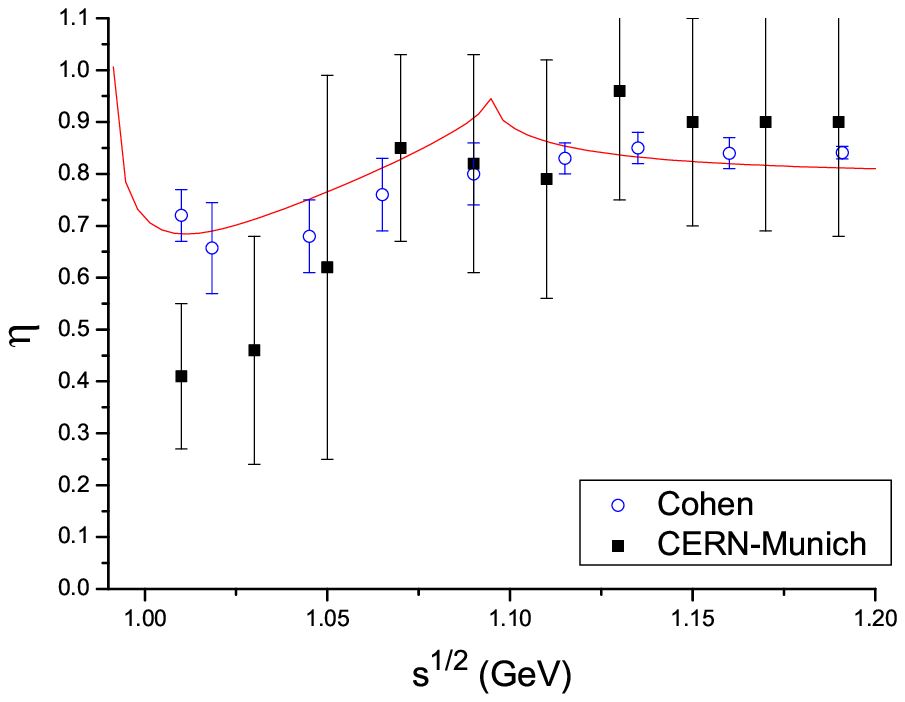}}%
\end{minipage}
\begin{minipage}[h]{0.5\textwidth}
\centering \subfigure[ ]{ \label{c}
\includegraphics[width=0.9\textwidth]{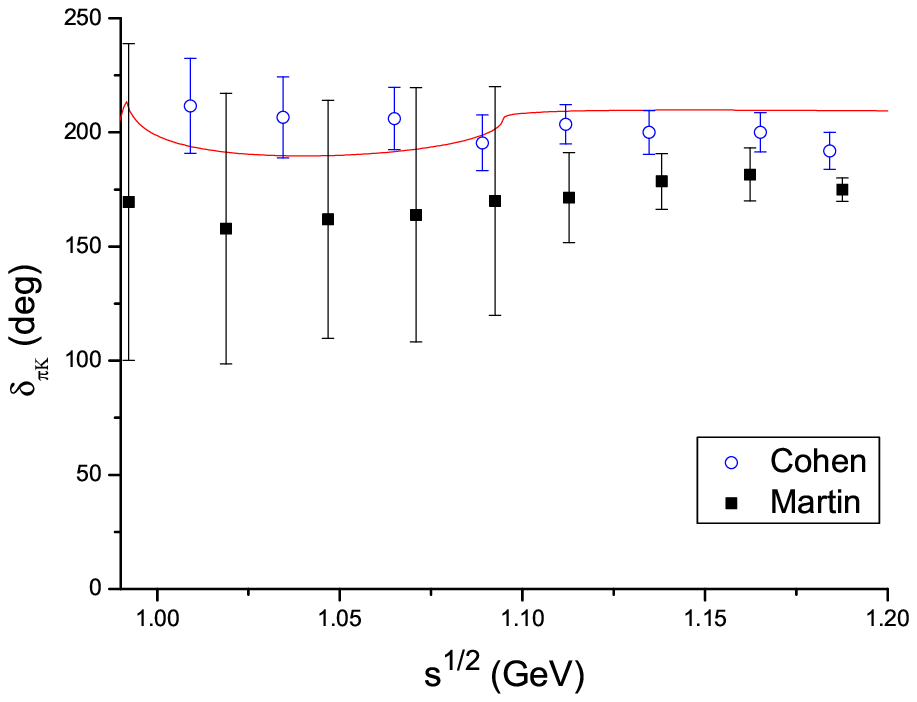}}%
\end{minipage}
\begin{minipage}[h]{0.5\textwidth}
\centering \subfigure[ ]{ \label{d}
\includegraphics[width=0.9\textwidth]{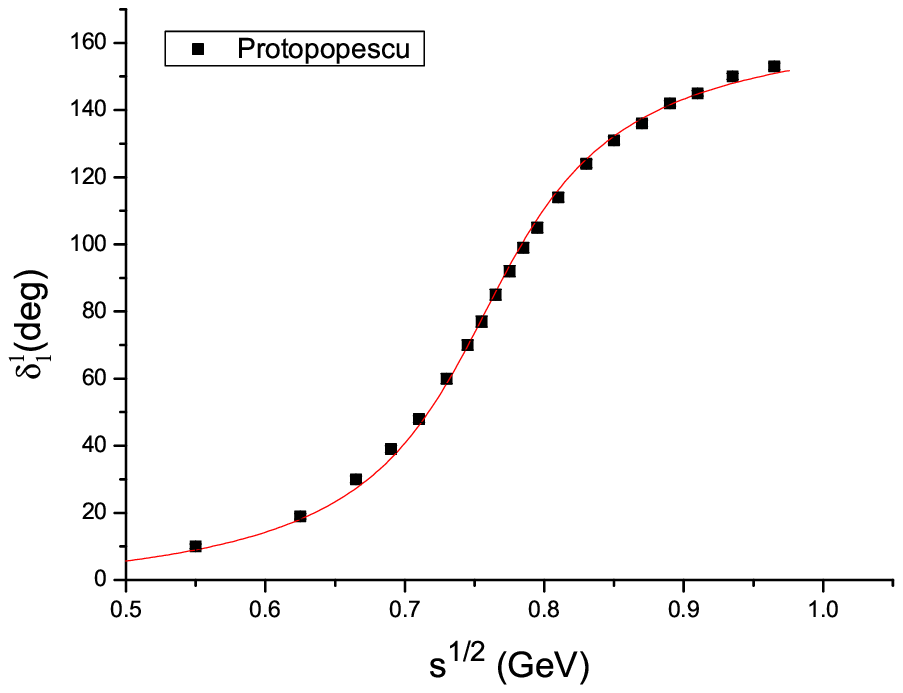}}%
\end{minipage}
\begin{minipage}[h]{0.5\textwidth}
\centering \subfigure[ ]{ \label{e}
\includegraphics[width=0.9\textwidth]{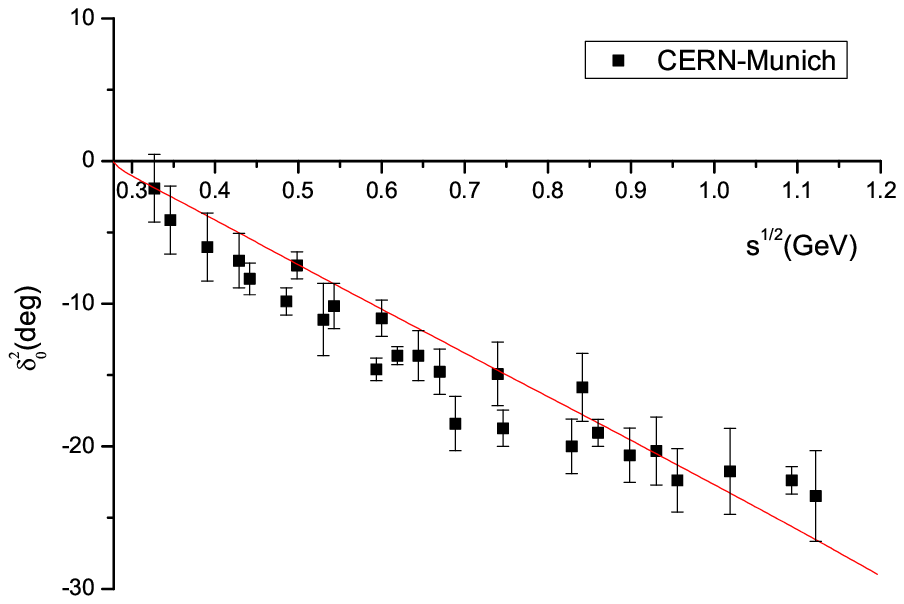}}%
\end{minipage}
\begin{minipage}[h]{0.5\textwidth}
\centering \subfigure[ ]{ \label{f}
\includegraphics[width=0.9\textwidth]{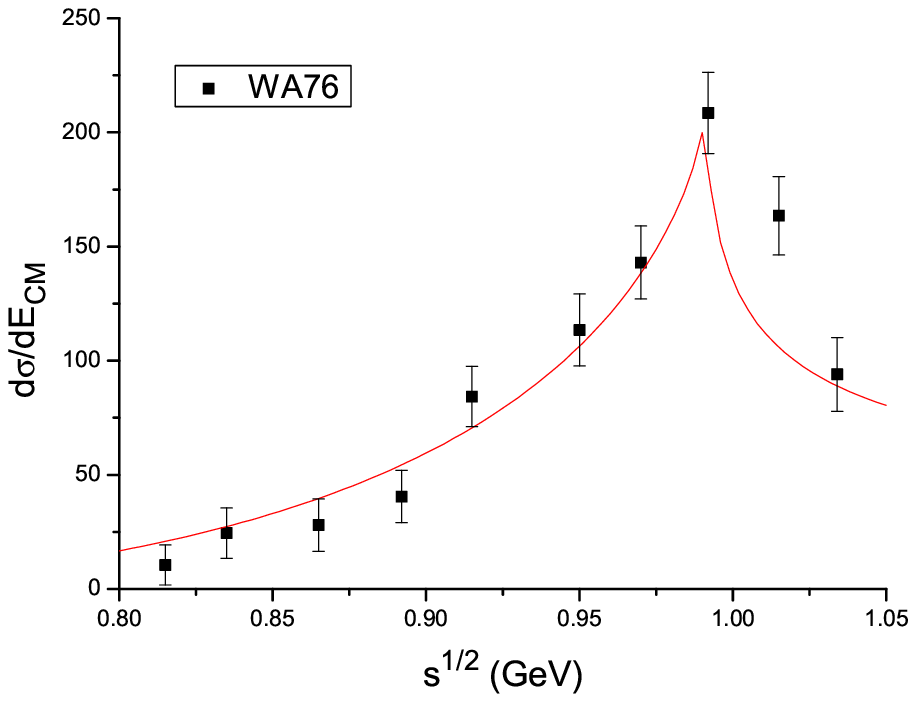}}
\end{minipage}
\caption{ \label{Fit2}The results of Fit 2.}
\end{figure}

\begin{figure}[h]%
\begin{minipage}[h]{0.5\textwidth}
\centering \subfigure[ ]{ \label{a}
\includegraphics[width=0.9\textwidth]{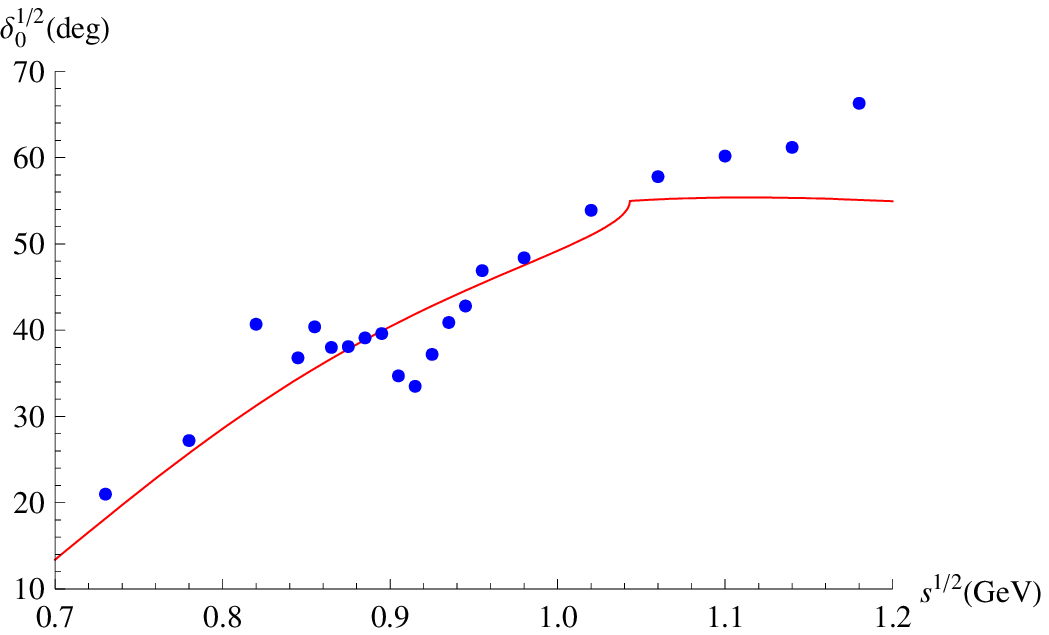}}%
\end{minipage}
\begin{minipage}[h]{0.5\textwidth}
\centering \subfigure[ ]{ \label{b}
\includegraphics[width=0.9\textwidth]{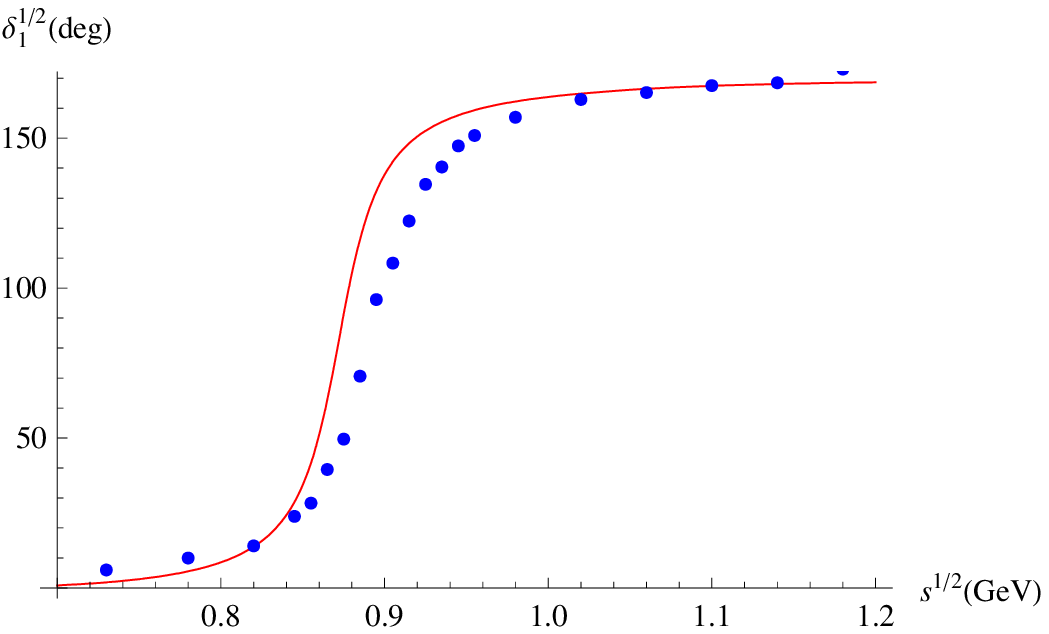}}%
\end{minipage}
\begin{minipage}[h]{0.5\textwidth}
\centering \subfigure[ ]{ \label{a}
\includegraphics[width=0.9\textwidth]{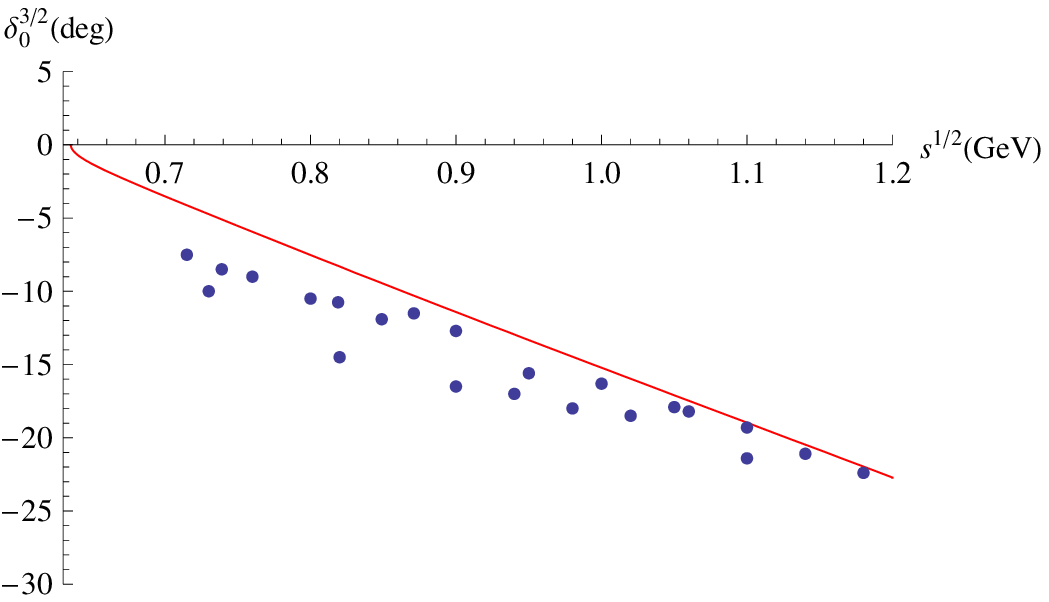}}%
\end{minipage}
\caption{ \label{Fit2-prediction}Predictions on phase shift from Fit
2. The data comes from~\cite{Estabrooks78,Linglin1973}}
\end{figure}

\begin{table}
\begin{center}
 \begin{tabular}  {|c|c|c|c|c|}\hline
 Channel &                &               ChPT\                    &                ChPT\                 &   Experiment  \\
  (I,J)  &  Our results   & $\mathcal {O}(p^4)$~\cite{Meissner91}  & $\mathcal {O}(p^6)$~\cite{Bijnens98} & ~\cite{Nagels79} \\ \hline
 $(0,0)$      &    0.208  &         0.20                           &        $0.220\pm0.005$                 & $ 0.26\pm0.05$   \\
 $(1,1)$      &    0.036  &         0.037                          &        $0.038\pm0.002$                 & $ 0.038\pm0.002$ \\
 $(2,0)$      &   -0.039  &        -0.042                          &       $-0.042\pm0.010$                 & $-0.028\pm0.012$ \\
 $(1,0)$      &    0.004  &         0.007                          &            \                         &       \        \\
 \hline\hline
 $(\frac{1}{2},0)$ & 0.169    &         0.17                &            \     &    0.13 to 0.24
 \\
$(\frac{1}{2},1)$  &  0.039   &         0.014               &  \
& 0.017 to 0.018   \\
$(\frac{3}{2},0)$  & -0.058   &         -0.05               &      \
& -0.13 to -0.05 \\ \hline
 \end{tabular}
 \caption{\label{Fit2-aIJ}From  Fit II: scattering lengths, in unit of $m_{\pi}^{-1}$ for S-wave and $m_{\pi}^{-3}$ for P-wave.}
 \end{center}
\end{table}
  In Fit II we further include I,J=2,0 channel of
$\pi\pi\rightarrow\pi\pi$ \cite{CERN-Munich} and I,J=1,0 channel of
$\pi\eta\rightarrow K\bar{K}$~\cite{WA76}.   For the latter, we use
the $\pi\eta$ effective mass distribution data from the
$pp\rightarrow p(\eta\pi^+\pi^-)p$ reaction studied by WA76
Collaboration~\cite{WA76}, and the same formula as~\cite{Pelaez02},
\begin{equation}
\frac{d\sigma}{dE_{cm}}=cp_{\pi\eta}|T_{\pi\eta\rightarrow
K\bar{K}}^{I,J=1,0}|^2+background\ ,
\end{equation}
where $c$ is a normalization factor.   The fit results are shown in
Fig.~\ref{Fit2} and Tab.~\ref{LECS}, with
$\chi^2_{d.o.f}=482/133=3.6$.
%Our result is $c=213\pm13$, to be
%compared with~\cite{Flatte76}, where $c$ was taken from 73 to 165
%$\mu b/\mathrm{GeV}^2$.
The background is subtracted in Fig.~\ref{Fit2}
(f)~\cite{Pelaez02,Flatte76}.   The LECs and scattering lengths for
different channels are given by Tab.~\ref{Fit2-aIJ}, which are in
agreement with $\chi$PT~\cite{Meissner91,Bijnens98} and experimental
results~\cite{Nagels79}. Notice that by fitting meson-meson
scattering data, we can not determine $L_6^r$, $L_7^r$ and $L_8^r$
individually, since these three parameters enter into the scattering
amplitudes as the combination $2L_6^r+L_8^r$ and $2L_7^r+L_8^r$. In
the I,J=2,0    channel there is a virtual state. The pole position
is at $s_v=0.045m_{\pi}^2$ on the II-sheet. A quite similar result
has been given in~\cite{Zhou05}.
\begin{table}
\begin{center}
 \begin{tabular}  {|c|c|c|c|}\hline
 Resonance        &        II          &        III        &  IV  \\ \hline
 $\sigma$         &  $0.457-i0.242$    &         \         &   \  \\
 $f_0(980)$       &  $0.974-i0.025$    &         \         &   \  \\
 $a_0(980)$       &        \           &   $0.640-i0.002$  &  $1.131-i0.079$  \\
     \            &        \           &         \         &   \  \\
 $\rho(770)$      &  $0.740-i0.069$    &   $0.782-i0.056$  &   \  \\
 $(I,J)=(2,0)$    &  $0.045 m_{\pi}^2$ &         \         &   \
 \\
\hline\hline
$\kappa$(800)     &  $0.673-i0.254$    &         \         &   \  \\
$K^{*}(892)$      &  $0.895-i0.026$    &   $0.921-i0.021$  &   \  \\
\hline
 \end{tabular}
 \caption{\label{Fit2-pole}Resonance pole positions on $\sqrt{s}$ plane in unit of $\mathrm{GeV}$,
 and virtual pole position on s plane.}
 \end{center}
\end{table}

\begin{figure}[h]%
\begin{minipage}[h]{0.5\textwidth}
\centering \subfigure[ ]{ \label{a}
\includegraphics[width=0.9\textwidth]{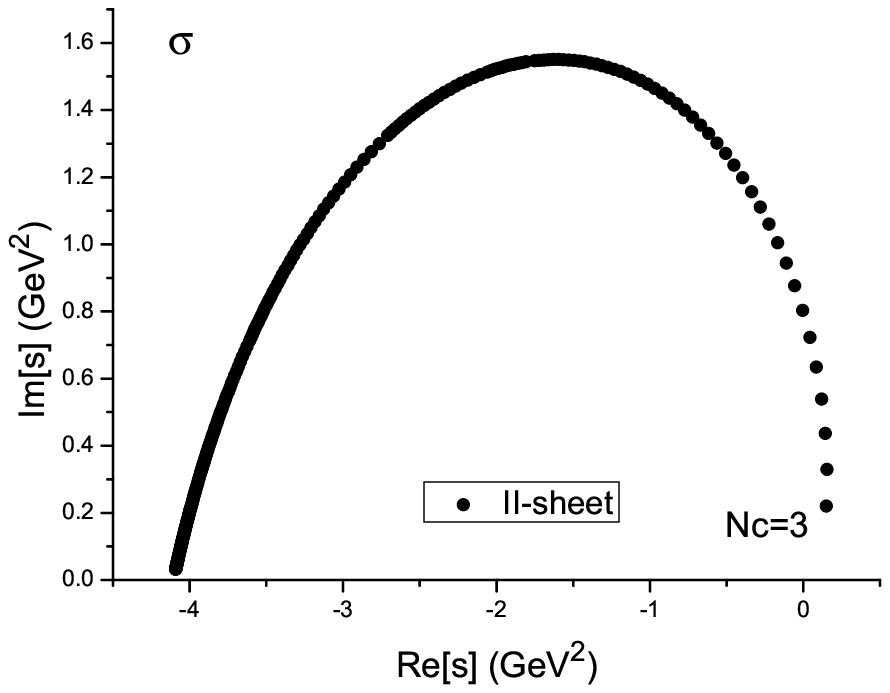}}%
\end{minipage}
\begin{minipage}[h]{0.5\textwidth}
\centering \subfigure[ ]{ \label{b}
\includegraphics[width=0.9\textwidth]{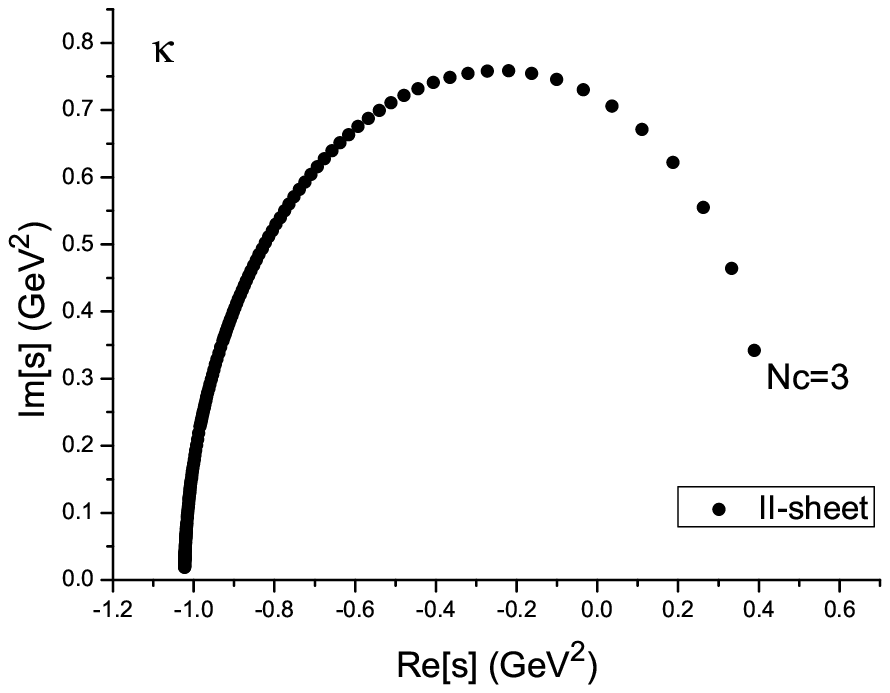}}%
\end{minipage}
\begin{minipage}[h]{0.5\textwidth}
\centering \subfigure[ ]{ \label{c}
\includegraphics[width=0.9\textwidth]{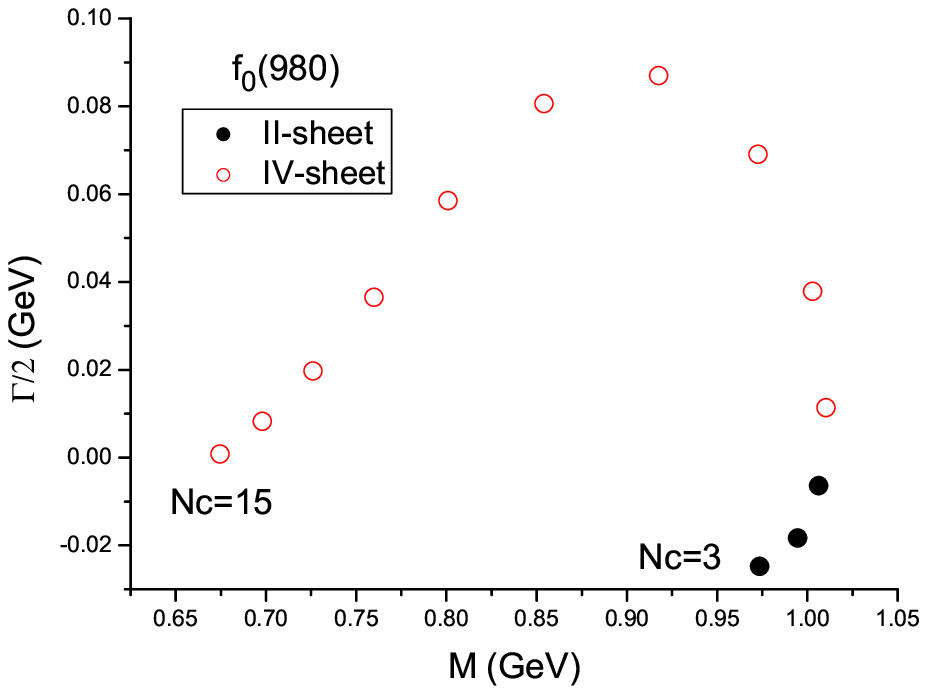}}%
\end{minipage}
\begin{minipage}[h]{0.5\textwidth}
\centering \subfigure[ ]{ \label{d}
\includegraphics[width=0.9\textwidth]{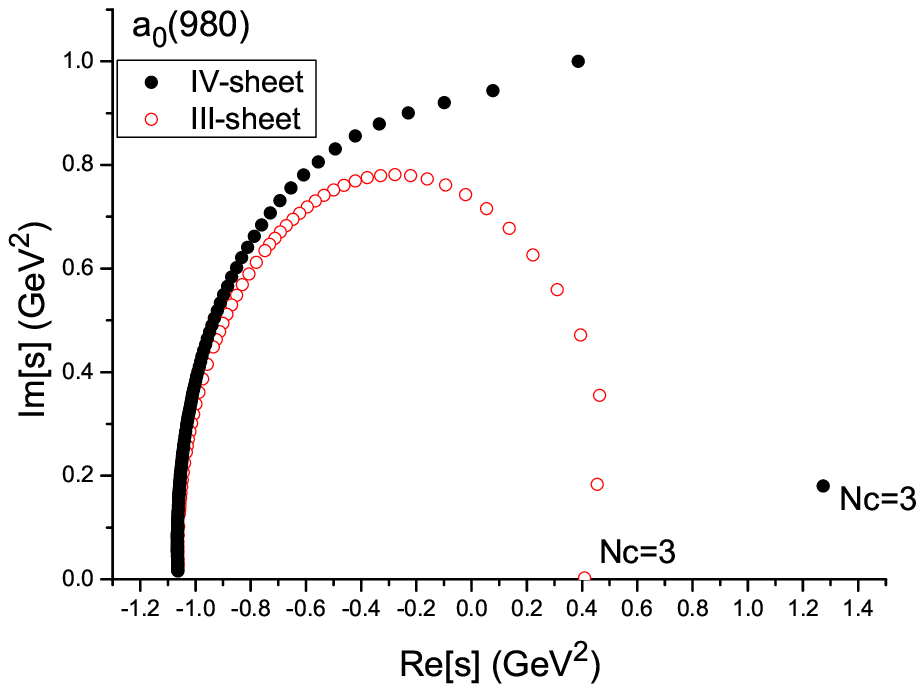}}%
\end{minipage}
\begin{minipage}[h]{0.5\textwidth}
\centering \subfigure[ ]{ \label{e}
\includegraphics[width=0.9\textwidth]{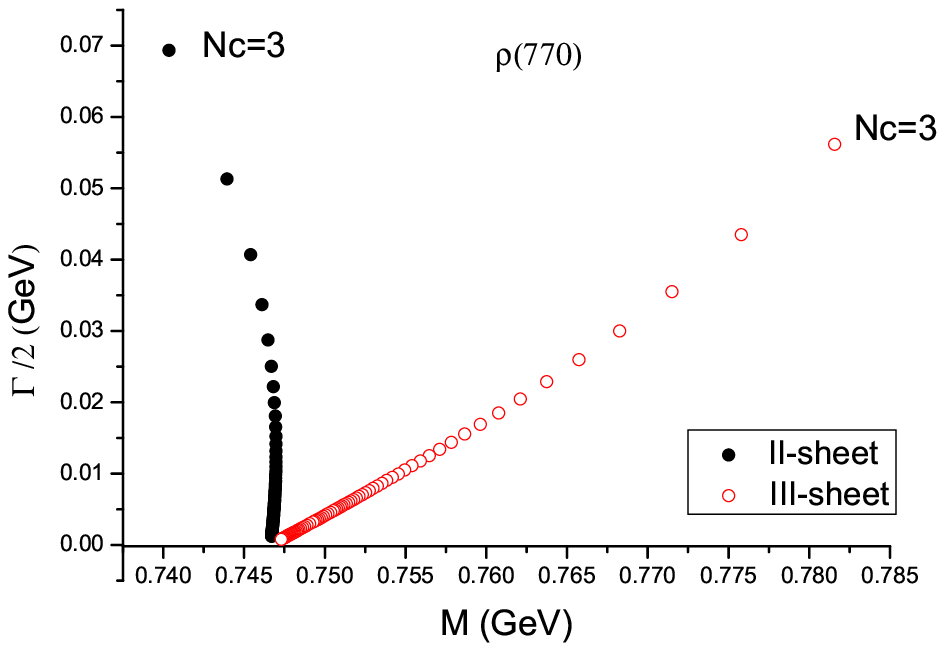}}%
\end{minipage}
\begin{minipage}[h]{0.5\textwidth}
\centering \subfigure[ ]{ \label{f}
\includegraphics[width=0.9\textwidth]{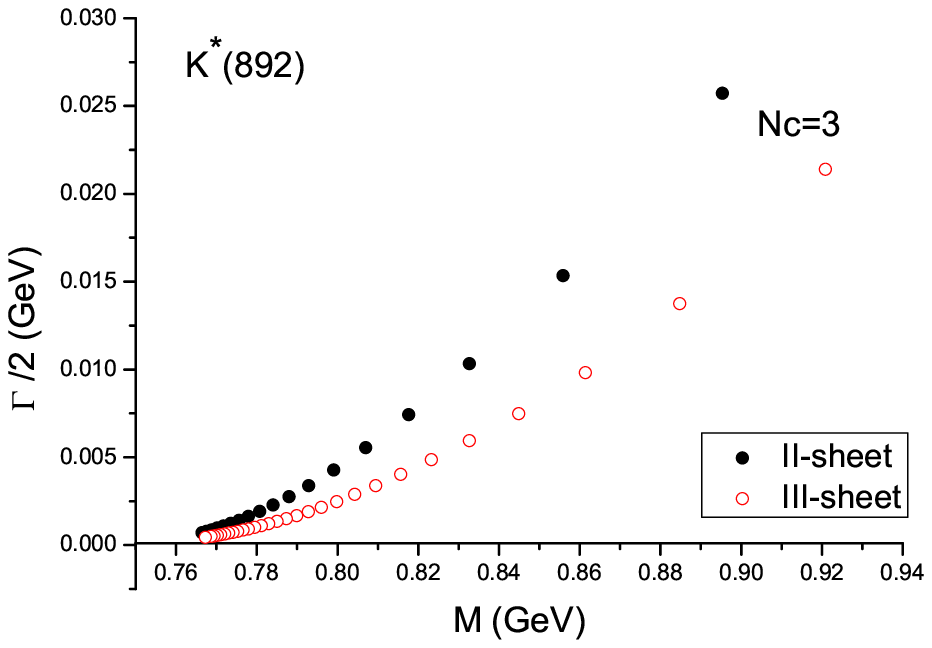}}
\end{minipage}
\caption{ \label{NctrajectoryFit2}$N_c$ trajectories of Fit 2.  }
\end{figure}
In this paper we do not fit the data in I,J=$\frac{1}{2},0$;
$\frac{1}{2},1$  and I,J=$\frac{3}{2},0$ $K\pi\rightarrow K\pi$
channels. Rather, we give predictions on the phase shifts,
scattering lengths in Fig.~\ref{Fit2-prediction} and
Tab.~\ref{Fit2-aIJ} respectively, which are in fairly good agreement
with the experimental results. We also give predictions on pole
positions of $K^*_0(800)$ and $K^*(892)$ when $N_c=3$, as well as
the trajectories when $N_c$ varies, as shown in
Fig.~\ref{NctrajectoryFit2}(b and f). In the following we briefly
summarize the major physical results.

Firstly, the results unambiguously indicate that $K^*(892)$ and
$\rho(770)$ are Breit-Wigner particles as they move to the real axis
of the complex $\sqrt{s}$-plane, on sheet II, where they meet their  shadow
partners on sheet III. Their masses and widths behave as $\mathcal
{O}(1)$ and $\mathcal {O}(1/N_c)$ respectively in large $N_c$ limit,
as expected for normal $q\bar{q}$ states. It is firstly observed in
~\cite{Pelaez03} that $\rho$ and $K^*$ behave as conventional $\bar
qq$ states in the unitarized amplitude since they fall down to the
real axis straightforwardly. But the twin pole structure of these
two particles was not discussed before. Hence the finding made in
the present paper certainly put the previous observation on a more
solid ground. This progress is important in the sense that  it
demonstrates  that Pad\'e approximants do faithfully reproduce the
correct analytical structure of a Breit--Wigner resonance.

 The $N_c$ trajectory of $\sigma(600)$ is quite similar to what we find in Fit
 1, and
 $K^*_0(800)$ has very similar trajectory comparing with that of
$f_0(600)$, not to mention that they both have low masses and broad
widths.
 We have made careful numerical
tests and did not find shadow poles related to these two poles. This
means that $\sigma$ and $\kappa$ basically only involve in  single
channel scattering.

The $N_c$ trajectory of $f_0(980)$ is also very similar to that in
Fit I. Again, careful numerical analysis is made and we find no
shadow pole associated with it.

  For $a_0(980)$, see table~\ref{Fit2-pole}, we find a sheet IV pole which is not too far away
from $K\bar{K}$ threshold. located at $\sqrt{s}=1.131-i0.079$MeV,
Its location is not quite the same as the experimentally  observed
$a_0(980)$, but we do not take this discrepancy seriously since it
may well be ascribed to the lack of enough good data in this
channel.
 What is truly surprising is that, drastically different
from $f_0(980)$, we find a companion shadow pole for $a_0(980)$ on
sheet III. Although far away from the physical region when $N_c=3$,
as shown in Fig.~\ref{NctrajectoryFit2}(d), both of them will move
to negative real axis, on sheet IV (or sheet III). Large $N_c$
techniques shows its powerfulness here as it can  pinpoint the same
origin of two totally different looking poles.\footnote{The pole
counting mechanism no longer works well for the present situation
since we encounter a strongly coupled system while the original pole
counting argument requires a weakly coupled CDD
pole~\cite{morgan92}. } The twin pole structure uncovers the nature
of $a_0(980)$ as a Breit--Wigner resonance.~\footnote{Nevertheless
falling down to the negative real axis on sheet IV makes it
difficult to be recognized. Here we remind again that the situation
may be like the $\sigma$ pole, i.e., its pole destination may be
parameter dependent.}

%%%%%%%%%%%%%%%%%%%%%%%%%%%%%%%%%%%%%%%%%%%%%%%%%%%%%%%%%%%%%%%%%%%%%%%%%%%%%%%%%%%

%%%%%%%%%%%%%%%%%%%%%%%%%%%%%%%%%%%%%%%%%%%%%%%%%%%%%%%%%%%%%%%%%%%%%%%%%%%%%%%%%%%%%%%%%%%%%%%%%%%%%%%%%%%
\section{Discussions and Conclusions}\label{Conclusions}

In this paper we have made efforts in exploring physics beyond the
the validity domain of $\chi$PT, by studying [1,1] matrix Pad\'e
amplitudes.  Masses and widths of resonances in all channels except
$a_0(980)$ are very well reproduced.
%There has actually been a
%rather long history that people work along this direction.
%It is
%know that one has to be very cautious to make predictions from
%Pad\'e amplitudes. For this purpose,
We propose to study the shadow pole structure of a resonance, which
is not emphasized in previous literature. We believe that this
method enables us to avoid as much as possible model dependence. It
is very encouraging to see, for the first time in the literature,
that Pad\'e approximants do faithfully exhibit the twin pole
structure of $\rho(770)$ and $K^*(892)$  -- particles widely
accepted as standard Breit--Wigner resonances, or $\bar qq$ states.
This, in our point view, also shed lights and strengthen the level
of confidence on the study of scalar resonances within Pad\'e
unitarization approach.

We trace numerically all poles' trajectories when $N_c$ varies. We
find that $\sigma$ and $\kappa$ have quite similar $N_c$
trajectories. They will move far away from the real axis on complex
s-plan at first, but eventually fall down to  real axis when $N_c$
is large enough. This confirms the previous results on
$\sigma$~\cite{Sun} and for $\kappa$ the present result is quite
new. The present result on $\sigma$ also agrees with that of
\cite{Guo11}, but for $\kappa$ pole the result is quite different.
The absence of shadow pole suggests that the $\sigma$ resonance is a
pure $SU(2)$ particle, i.e., does not contain $\bar ss$ component.

 The
result of $f_0(980)$ is also interesting. The sheet II pole will
move to the real axis of IV-sheet above $\pi\pi$ threshold, winding
around the branching point of $K\bar{K}$ threshold. This suggests
that $f_0(980)$ may be regarded as a molecular state in reality and
gradually becomes a virtual bound state when $N_c$ becomes large.

It is awesome to see that a twin pole structure of $a_0(980)$ is
also found even though it behaves quite oddly when $N_c$ becomes
large. If the picture shown in this paper is correct then the
$a_0(980)$ may be explained as a couple channel Breit--Wigner or
$\bar qq$ state. Notice that the conclusion reached in this paper is
compatible with the picture drawn in Ref.~\cite{kala} where it is
found that $a_0(980)$ is more `elementary' than $f_0(980)$. Also the
present result is compatible with the result of Ref.~\cite{xiaoly}
where it is suggested that it is not possible to put $f_0(980)$ into
a $\bar qq$ octet.

Predictions from Pad\'e approximants may be model and parameter dependent. Its pole structure, on the contrary,
is expected to be immune from these deficiencies.
We hope that the exploration made in this this paper would stimulate
more fruitful studies in future in exploring the properties of light
scalars which is certainly interesting and challenging.

%%%%%%%%%%%%%%%%%%%%%%%%%%%%%%%%%%%%%%%%%%%%%%%%%%%%%%%%%%%%%%%%%%%%%%%%%%%%%%%%%%%%%%%%%%%%%%%
\section{Acknowledgement}
This work is supported in part by  National Nature Science
Foundations of China under contract number  10925522, 10875001 and
11021092.
%%%%%%%%%%%%%%%%%%%%%%%%%%%%%%%%%%%%%%%%%%%%%%%%%%%%%%%%%%%%%%%%%%%%%%%%%%%%%%%%%%%%%%%%%%%%%%%%%%%%%%%%%%%%%


\begin{thebibliography}{99}
\bibitem{Weinberg79}S. Weinberg, Physica {\bf A96} (1979) 327.
\bibitem{Gasser84}J.~Gasser, H.~Leutwyler, Ann. Phys. (NY) 158 (1984) 142.
\bibitem{Gasser85}J.~Gasser, H.~Leutwyler, Nucl. Phys. {\bf B250} (1985) 465.

\bibitem{Zheng01}Z.~G.~Xiao, H.~Q.~Zheng, Nucl. Phys. {\bf A695} (2001) 273.

\bibitem{Zheng04}Z.~Y. Zhou, H.~Q.~Zheng,  Nucl. Phys. {\bf A775} (2006) 212;
 H.~Q.~Zheng, \textit{et al.}, Nucl. Phys. {\bf A733} (2004) 235.

\bibitem{Zhou05}Z.~Y.~Zhou \textit{et al.}, J. High Energy Phys. 02 (2005) 043.

\bibitem{CGL}I.~Caprini, G.~Colangelo, H.~Leutwyler, Phys. Rev. Lett. {\bf 96} (2006) 132001.

\bibitem{Moussallam06}S.~Descotes-Genon, B.~Moussallam, Eur. Phys. J. {\bf C48} (2006) 553.

\bibitem{KPY}R.~Kaminski, J.~R.~Pelaez, F.~J.~Yndurain, Phys. Rev. {\bf D77} (2008) 054015.

\bibitem{Pennington06}M.~R.~Pennington, Invited talk at YKIS Seminaron New Frontiers in QCD: Exotic Hadrons and Hadronic Matter, Kyoto,
Japan, 20 Nov-8 Dec 2006. Prog. Theor. Phys. Suppl. 168 (2007) 143.

\bibitem{Achasov08}N.~N.~Achasov, G.~N.~Shestakov, Phys. Rev. {\bf D77} (2008) 074020.

\bibitem{MNO08}G.~Mennessier, S.~Narison, W.~Ochs, Phys. Lett. {\bf B665} (2008) 205.

\bibitem{YuMao}Yu.~Mao, \textit{et al.}, Phys. Rev. {\bf D79} (2009) 116008.

%\bibitem{MNW1}G.~Mennessier, S.~Narison, X.~G.~Wang, Phys. Lett. {\bf B688} (2010) 59.

\bibitem{MNW2} G.~Mennessier, S.~Narison, X.~G.~Wang, Phys. Lett. {\bf B688} (2010) 59;
 G.~Mennessier, S.~Narison, X.~G.~Wang, Phys. Lett. {\bf B696} (2011) 40.

\bibitem{Pelaez03}J.~R.~Pelaez, arXiv:0306063[hep-ph]; AIP Conf. Proc. 687 (2003) 74.
\bibitem{Pelaez04}J.~R.~Pelaez, Phys. Rev. Lett. 92 (2004) 102001.
\bibitem{Pelaez06}J.~R.~Pelaez and G.~Rios, Phys. Rev. Lett. 97 (2006) 242002.
\bibitem{Sun}Z.~X.~Sun \textit{et al.}, Mod. Phys. Lett.{\bf A22} (2007) 711.
\bibitem{Uehara04}M.~Uehara, hep-ph/0404221.
\bibitem{Guo11}Z.~H.~Guo, J.~A.~Oller, arXiv:1104.2849[hep-ph].

\bibitem{Pelaez02}A.~G.~Nicola, J.~R.~Pelaez, Phys. Rev. {\bf D65} (2002) 054009.

\bibitem{Kennedy62}J.~Kennedy, T.~D.~Spearman, Phys. Rev.{\bf 126} (1962) 1596.
\bibitem{Benitez78}M.~Aguilar-Benitez \textit{et al.}, Nucl. Phys. {\bf B140} (1978) 73.

\bibitem{Oller99}F.~Guerrero, J.~A.~Oller, Nucl. Phys. {\bf B537} (1999) 459.

\bibitem{CERN-Munich}W.~Ochs, Ph.D. thesis, Munich Univ., 1974.
\bibitem{NA48}J.~R.~Batley \textit{et al.}, Eur. Phys. J. {\bf C52} (2007) 875.
\bibitem{Cohen}D.~Cohen \textit{et al.}, Phys. Rev. {\bf D22} (1980) 2595.
\bibitem{Martin}A.~D.~Martin and E.~N.~Ozmultu, Nucl. Phys. {\bf B158} (1979) 5201.
\bibitem{Protopopescu}S.~D.~Protopopescu and M.~Alson-Granjost, Phys. Rev. {\bf D7} (1973) 1279.

\bibitem{Cillero}Z.~H.~Guo, J.~J.~Sanz Cillero, H.~Q.~Zheng, JHEP {\bf 0706} (2007) 030.

\bibitem{morgan92}D.~Morgan, Nucl. Phys. {\bf A543} (1992) 632.

\bibitem{Zhangou} Ou Zhang, C. Meng, H.~Q.~Zheng,  Phys. Lett. {\bf B680} (2009) 453.
%\bibitem{Jaffe81}R.~L.~Jaffe, Proceedings of the Intl.Symposium on
%Lepton and Photon Interactions at High Energies. Physikalisches
%Institut, University of Bonn(1981). ISBN:3-9800625-0-3.

\bibitem{WA76}T.A.~Armstrong \textit{et al.}, Z. Phys. {\bf C52} (1991) 389.
\bibitem{Flatte76}S.M.~Flatt\'{e}, Phys. Lett. {\bf B63} (1976) 224.

\bibitem{Meissner91}V.~Bernard, N.~Kaiser, U.~G.~Meissner, Nucl. Phys. {\bf B357}(1991)129; Nucl. Phys. {\bf B364} (1991) 283.
\bibitem{Bijnens98}J.~Bijnens, G.~Colangelo and P.~Talavera, J. High Energy Phys. {\bf 05} (1998) 014.
\bibitem{Nagels79}M.M.~Nagels \textit{et al.}, Nucl. Phys. {\bf B147} (1979) 189.

\bibitem{Estabrooks78}P. Estabrooks \textit{et al.}, Nucl. Phys. {\bf B133} (1978) 490.
\bibitem{Linglin1973}D.~Linglin \textit{et al.}, Nucl. Phys. {\bf B57} (1973) 64.

\bibitem{kala}V.~Baru,  J.~Haidenbauer, C.~Hanhart,  Yu.~Kalashnikova, A.~Kudryavtsev,
Phys. Lett. {\bf B586} (2004) 53.

\bibitem{xiaoly} M.~X.~Su, L.~Y.~Xiao, H.~Q.~Zheng, Nucl. Phys. {\bf A792} (2007) 288.

%\bibitem{Baker1975}S.~L.~Baker \textit{et al.}, Nucl. Phys. {\bf B99} (1975) 211.
%\bibitem{Mercer1971}R.~Mercer \textit{et al.}, Nucl. Phys. {\bf B32} (1971) 381.


%\bibitem{Xiao}Z.G.~Xiao and H.Q.~Zheng, arXiv:0502199[hep-ph].

\end{thebibliography}
\end{document}